\documentclass[a4paper,fleqn,usenatbib]{mnras}

\usepackage{newtxtext,newtxmath}

\usepackage[T1]{fontenc}
\usepackage{ae,aecompl}

\usepackage{graphicx}	
\usepackage{amsmath}	
\usepackage{amssymb}	
\usepackage{threeparttable}
\usepackage{booktabs}

\def \lya{Ly$\alpha$}

\def \h2{{\rm H_{2}}}

\def \dex{{\rm dex}}

\def \Cii{[\ion{C}{ii}]}

\def \oiii{[\ion{O}{iii}]}
\def \nii{[\ion{N}{ii}]}

\def \LIR{L_{{\rm IR}}}
\def \LFIR{L_{{\rm FIR}}}

\def \IRXB{IRX$-\beta$}
\def \dn4000{D_{{\rm n}}(4000) }

\title[Dust Temperature of $z\sim5.5$ Galaxies]{ALMA Characterises the Dust Temperature of $z\sim5.5$ Star-Forming Galaxies}

\author[A. L. Faisst et al.]{
Andreas L. Faisst$^{1}$\thanks{E-mail: afaisst@caltech.edu},
Yoshinobu Fudamoto$^{2}$,
Pascal A. Oesch$^{2,3}$,
Nick Scoville$^{4}$,\newauthor
\,\,Dominik A. Riechers$^{5,6}$,
Riccardo Pavesi$^{5}$
and Peter Capak$^{3,4}$\newauthor
\\
$^{1}$IPAC, California Institute of Technology
1200 E California Boulevard, Pasadena, CA 91125, USA\\
$^{2}$Observatoire de Gen\`eve, Universit\'e de Gen\`eve, 51 Ch. des Maillettes, 1290 Versoix, Switzerland\\
$^{3}$Cosmic Dawn Center (DAWN) at the Niels Bohr Institute, University of Copenhagen and National Space Institute, Technical University of Denmark\\
$^{4}$California Institute of Technology, MC 249-17, 1200 East California Boulevard, Pasadena, CA 91125, USA\\
$^{5}$Department of Astronomy, Cornell University, Space Sciences Building, Ithaca, NY 14853, USA\\
$^{6}$Max-Planck-Institut f\"ur Astronomie, K\"onigstuhl 17, D-69117 Heidelberg, Germany
}

\date{Accepted XXX. Received YYY; in original form ZZZ}

\pubyear{2020}

\begin{document}
\label{firstpage}
\pagerange{\pageref{firstpage}--\pageref{lastpage}}
\maketitle

\begin{abstract}
The infrared spectral energy distributions (SEDs) of main-sequence galaxies in the early universe ($z>4$) is currently unconstrained as infrared continuum observations are time consuming and not feasible for large samples.
We present Atacama Large Millimetre Array (ALMA) Band 8 observations of four main-sequence galaxies at $z\sim5.5$ to study their infrared SED shape in detail.
Our continuum data (rest-frame 110$\rm \mu m$, close to the peak of infrared emission) allows us to constrain luminosity weighted dust temperatures and total infrared luminosities. With data at longer wavelengths, we measure for the first time the emissivity index at these redshifts to provide more robust estimates of molecular gas masses based on dust continuum.
The Band 8 observations of three out of four galaxies can only be reconciled with optically thin emission redward of rest-frame $100\,{\rm \mu m}$. The derived dust peak temperatures at $z\sim5.5$ ($38\pm8\,{\rm K}$) are elevated compared to average local galaxies, however, $5-10\,{\rm K}$ below what would be predicted from an extrapolation of the trend at $z<4$. This behaviour can be explained by decreasing dust abundance (or density) towards high redshifts, which would cause the infrared SED at the peak to be more optically thin, making hot dust more visible to the external observer. From the $850{\rm \mu m}$ dust continuum, we derive molecular gas masses between $10^{10}$ and $10^{11}\,{\rm M_{\odot}}$ and gas fractions (gas over total mass) of $30-80\%$ (gas depletion times of $100-220\,{\rm Myrs}$).
All in all, our results provide a first measured benchmark SED to interpret future millimetre observations of normal, main-sequence galaxies in the early Universe.
\end{abstract}




\begin{table*}
\caption{Summary of Observations.} \label{tab:observations}
\begin{tabular}{l | lcc | lcc | lcc }
  \hline\hline
  ID & \multicolumn{3}{c}{Band 6} & \multicolumn{3}{c}{Band 7}  &  \multicolumn{3}{c}{Band 8} \\
  \cmidrule(lr){2-4} \cmidrule(lr){5-7}  \cmidrule(lr){8-10}
    & PID & Resolution & $\sigma$ & PID & Resolution & $\sigma$ & PID & Resolution & $\sigma$ \\
     &  &  &[$\rm{\mu Jy}$] &  &  & [$\rm{\mu Jy}$] &  &  & [$\rm{\mu Jy}$] \\ \hline
        {\it HZ4} & 2015.1.00388.S$^1$ & 1.1\arcsec & 14 & 2017.1.00428.L$^3$ & 0.93\arcsec & 21 & 2018.1.00348.S$^5$ & 0.71\arcsec & 34\\
        {\it HZ6} & 2015.1.00388.S$^1$, 2015.1.00928.S$^2$ & 1.4\arcsec & 23 & 2017.1.00428.L$^3$ & 0.89\arcsec & 29 & 2018.1.00348.S$^5$ & 0.73\arcsec & 40\\
        {\it HZ9} & 2015.1.00388.S$^1$ & 1.4\arcsec & 14 & 2012.1.00523.S$^4$ & 0.58\arcsec & 41 & 2018.1.00348.S$^5$ & 0.73\arcsec & 57\\
        {\it HZ10} & 2015.1.00388.S$^1$, 2015.1.00928.S$^2$ & 1.2\arcsec & 21 & 2012.1.00523.S$^4$ & 0.58\arcsec & 53 & 2018.1.00348.S$^5$ & 0.68\arcsec & 66 \\ \hline
 \end{tabular}\\
  \begin{tablenotes}
   \item \textbf{Notes: } The quoted $\sigma$ represents the RMS of the continuum images per beam.  References: (1) \citet[][]{LU18}, (2) \citet[][]{PAVESI19}, (3) \citet[][]{ALPINE_BETHERMIN19}, (4) \citet[][]{CAPAK15}, (5) this work.
  \end{tablenotes}
 \end{table*}


\section{Introduction} \label{sec:intro}

Galaxies evolve significantly during the first $1-2\,{\rm Gyrs}$ after the Big Bang. Specifically, after the Epoch of Reioinsation at redshifts $4~<~z~<~6$, galaxies establish fundamental properties as they transition from a primordial to a more mature state. For example, altered optical line ratios are consistent with a harder ionising radiation field in early galaxies and/or a changing configuration of molecular clouds from density to radiation bounded \citep[e.g.,][]{LABBE13,DEBARROS14,NAKAJIMA14,FAISST16c,HARIKANE19}.
Connected to this, the average metal content of galaxies is increasing from sub-solar to solar during this time \citep[][]{ANDO07,MANNUCCI10,FAISST16b}.
Going along with the metal enrichment is the rapid growth in stellar mass through mergers and the accretion of pristine gas \citep{BOUCHE12,LILLY13,FAISST16a,DAVIDZON17,SCOVILLE17,DAVIDZON18}.
Finally, the ultra-violet (UV) colours of galaxies at high redshifts tend to be bluer compared to their descendants, which is indicative of less reddening of their UV light due to dust \citep[e.g.,][]{BOUWENS09,BOUWENS12,FINKELSTEIN12}.

The \textit{Atacama Large (Sub-) Millimetre Array} (ALMA) has enabled us to extend these previous studies into the far-infrared (far-IR) light through observations of the far-IR continuum and emission lines, commonly the singly ionised Carbon atom (C$^{+}$, $158\,{\rm \mu m}$), in normal main-sequence galaxies at $>4$ \citep[e.g.,][]{WALTER12,WILLOTT15,RIECHERS14,CAPAK15}.
The recently completed \textit{ALMA Large Program to Investigate C$^+$ at Early Times} \citep[ALPINE,][]{ALPINE_LEFEVRE19,ALPINE_FAISST20,ALPINE_BETHERMIN19}\footnote{\url{http://alpine.ipac.caltech.edu}} provides such measurements for the largest sample of $z=4-6$ main-sequence galaxies to-date. \textit{ALPINE} builds the state-of-the-art for the characterisation of dust and gas in early galaxies in conjunction with the wealth of ancillary UV and optical datasets \citep[see also][]{FAISST19}.

From these ALMA observations, our understanding of the interstellar medium (ISM) of galaxies in the early universe has strongly progressed.
The evolution of the \IRXB~relation\footnote{It relates the ratio of rest-UV and total infrared luminosity to the rest-UV continuum slope $\beta$ \citep{MEURER99}.} with redshift has taught us about changes in dust attenuation.
While most galaxies at $z<4$ show similar dust attenuation properties as local starburst galaxies \citep[e.g.,][]{FUDAMOTO17}, recent studies based on the \textit{ALPINE} sample suggest a significant drop in dust attenuation at $z>4$ \citep{ALPINE_FUDAMOTO20} thereby approaching the dust properties of the metal-poor Small Magellanic Cloud \citep{PREVOT84}.
Furthermore, the total infrared luminosity is crucial to derive total star formation rates \citep[SFR,][]{KENNICUTT98} that tell about the true growth rates of galaxies at high redshifts and the evolution of the main-sequence with cosmic time \citep{ALPINE_KHUSANOVA20}.
Finally, the far-IR dust continuum emitted in the optically thin Rayleigh-Jeans (RJ) part of the far-IR spectral energy distribution (SED) at $>250\,{\rm \mu m}$ has turned out to be a good proxy of the total molecular gas mass of a galaxy \citep[e.g.,][]{SCOVILLE14}. This alternative method is crucial as deriving gas masses directly from observations of CO transitions is time consuming at these redshifts. Studies of large samples of galaxies with far-IR continuum measurements up to $z\sim6$ provide important constraints on the evolution of molecular gas and help us to understand how these galaxies form \citep[][]{SCOVILLE16,KAASINEN19,ALPINE_DESSAUGES20}.

However, the robustness of the results mentioned above is significantly limited by the fact that the infrared SED is inherently unknown at high redshifts \citep[see][]{FAISST17b}.
The relative faintness of these galaxies makes infrared continuum measurements time consuming and they are often secondary and only pursued in parallel with the observation of strong far-IR emission lines such as C$^+$, \nii, or \oiii.
The measurement of all infrared quantities (total luminosities, SFRs, molecular gas masses, etc) are therefore significantly relying on assumptions on the shape of the infrared SED. These assumptions are commonly based on SEDs of galaxies at lower redshifts.
The luminosity weighted temperature of the infrared SED is one of the key variables that define its shape. As shown in \citet[][]{FAISST17b}, using an average temperature based on low-redshift galaxies can underestimate the true total infrared luminosity by up to a factor of five. There is observational and theoretical evidence that galaxies at high redshifts are warmer \citep[e.g.,][]{MAGDIS12,MAGNELLI14,BETHERMIN15,FERRARA17,SCHREIBER18,LIANG19,MA19,SOMMOVIGO20}, which could be related to their lower metal content or higher star formation density. Such a relation is expected from studies of local galaxies \citep{FAISST17b}.
To characterise changes in the infrared SED of galaxies at $z>4$ to verify (or disprove) current assumptions, wavelengths closer to the peak of the infrared emission (around rest-frame $100\,{\rm \mu m}$) have to be probed. 

In this paper, we present new ALMA measurements at rest-frame $110\,{\rm \mu m}$ (Band 8) for four main-sequence galaxies at $z~\sim~5.5$. Note that Band 8 provides the strongest constraints on the location of the peak of the infrared SED (and hence luminosity weighted dust temperature) while minimising the observation time with ALMA.
These measurements are combined with archival data at rest-frame $150\,{\rm \mu m}$ (Band 7) and $205\,{\rm \mu m}$ (Band 6) to provide improved constraints on the infrared SEDs of high-redshift galaxies. A comparison to lower redshifts gives us important insights into the evolution of dust properties.

This paper is organised as follows: In Section~\ref{sec:observations}, we detail our new observations together with the archival data. In Section~\ref{sec:measurements}, we outline the procedure of fitting the infrared SEDs together with the measurements of dust temperature, total infrared luminosities, and molecular gas masses. We discuss the temperature$-$redshift evolution and a possible physical meaning using an analytical model in Section~\ref{sec:discussion} and conclude in Section~\ref{sec:end}. 
Throughout this work, we assume a $\Lambda$CDM cosmology with $H_0 = 70\,{\rm km\,s^{-1}\,Mpc^{-1}}$, $\Omega_\Lambda = 0.70$, and $\Omega_{\rm m} = 0.30$. All magnitudes are given in the AB system \citep{OKE74} and stellar masses and SFRs are normalised to a \citet[][]{CHABRIER03} initial mass function (IMF).

\begin{figure*}
    \centering
    \includegraphics[width=0.95\textwidth]{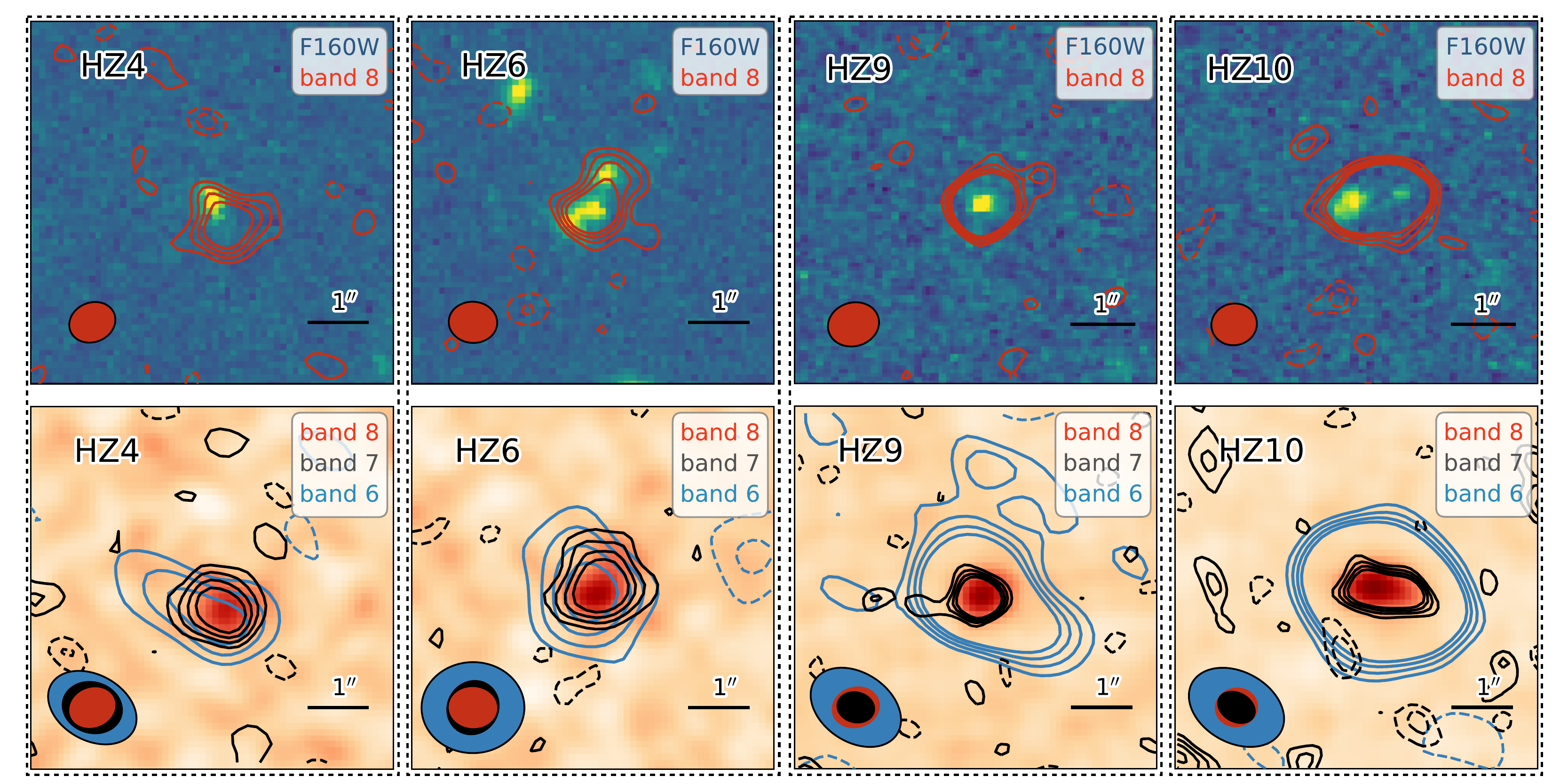}
    \caption{$6\arcsec\times6\arcsec$ cutouts of our four sources. North is up and east is to the left. \textit{Top Panels:} HST F160W (rest-frame $\sim2500\,\mathring{\rm A}$) cutouts with ALMA Band 8 continuum contours (red).
    \textit{Bottom Panels:} ALMA Band 8 continuum (background image) with contours showing Band 7 (black) and Band 6 (blue) continuum data from the literature.
    The solid contours start from 2$\,\sigma$ and end at $5\,\sigma$, and dashed contours indicate $-3\,\sigma$ and $-2\,\sigma$. The ALMA synthesised beams are shown for the different bands.
    All galaxies are significantly ($5-25\sigma$) detected in the far-IR continuum, and their flux peak positions are consistent each other within the beam sizes. There is a potential offset of the positions between the HST image and the far-IR detection for \textit{HZ4}. 
    }
    \label{fig:cutouts}
\end{figure*}

\begin{table*}
\caption{Summary of Flux Measurements.} \label{tab:measurements}
\begin{tabular}{lccccccc}
  \hline\hline
  ID & z & $\lambda_{\rm band6}$ & $f_{\rm{band6}}$ & $\lambda_{\rm band7}$ & $f_{\rm{band7}}$ & $\lambda_{\rm band8}$ & $f_{\rm{band8}}$\\
  &   & $\left[\rm{\mu m}\right]$ & $\rm{\left[\mu Jy\right]}$ & $\left[\rm{\mu m}\right]$ & $\rm{\left[\mu Jy\right]}$ & $\left[\rm{\mu m}\right]$ & $\rm{\left[\mu Jy\right]}$ \\ \hline
   {\it HZ4} & 5.544 & 1294 & $102\pm 26\,(\pm6)$ & 1014 & $189\pm 30\,(\pm9)$ & 738 & $524\pm 88\,(\pm31)$ \\
        {\it HZ6} & 5.293 & 1328 & $256\pm55\,(\pm23)$ & 975 & $404\pm61\,(\pm18)$ & 738 & $610\pm86\,(\pm37)$ \\
        {\it HZ9} & 5.541 & 1294 &  $274\pm22\,(\pm18)$ & 1008 &  $570\pm67\,(\pm29)$ & 738 & $1109\pm84\,(\pm67)$ \\
        {\it HZ10} & 5.657 & 1318 & $706\pm25\,(\pm35)$ & 1027 & $1519\pm74\,(\pm76)$ & 738 & $2813\pm129\,(\pm169)$ \\ \hline
 \end{tabular}\\
  \begin{tablenotes}
   \item \textbf{Notes: } Flux errors in the parentheses are calibration error estimated in \S\ref{sec:calibrationerror}.
  \end{tablenotes}
\end{table*}


\section{Data}
\label{sec:observations}

We focus on four main-sequence galaxies to which we refer to as \textit{HZ4} ($z=5.544$), \textit{HZ6} ($z=5.293$)\footnote{This galaxy is also known as \textit{LBG-1} in \citet[][]{RIECHERS14}.}, \textit{HZ9} ($z=5.541$), and \textit{HZ10} ($z=5.657$) in the following. These galaxies have been previously discussed by \citet[][]{RIECHERS14} and \citet[][]{CAPAK15} and are initially spectroscopically selected via \lya~and UV absorption lines from a large spectroscopic campaign with Keck/DEIMOS \citep{HASINGER18} on the \textit{Cosmic Evolution Survey} \citep[COSMOS,][]{SCOVILLE07} field.
All galaxies have been observed with different ALMA programmes covering their rest-frame wavelengths from $100\,{\rm \mu m}$ to $200\,{\rm \mu m}$ (see Table~\ref{tab:observations}).

\subsection{New ALMA Band 8 Observation}
\label{sec:band8}

All four galaxies have been observed recently as part of the ALMA programme \#2018.1.00348.S (PI: Faisst) at a frequency of $406.4\,{\rm GHz}$ (Band 8). This frequency was chosen to optimise the constraints on the infrared SED and to minimise the integration time to reach a S/N of $10$. At that frequency, the Band 8 atmospheric transmission is maximised and going to higher frequencies would increase the integration times significantly. On the other hand, Band 7 observes too low frequencies for robust constraints on the infrared SED together with the archival ALMA observations \citep[see Appendix in][]{FAISST17b}.
In the rest-frame of \textit{HZ4}, \textit{HZ6}, \textit{HZ9}, and \textit{HZ10}, Band 8 corresponds to wavelengths of $112.8\,{\rm \mu m}$,
$117.3\,{\rm \mu m}$, $112.9\,{\rm \mu m}$, and $111.0\,{\rm \mu m}$, respectively.
The observations were carried out in Cycle 6 between January 9 and 12, 2019 in the C43-2 compact configuration (maximal baseline $\sim300\,{\rm m}$) at an angular resolution of $0.55\arcsec$ to $0.62\arcsec$ under good weather conditions (precipitable water vapor column between $0.38\,{\rm mm}$ and $0.87\,{\rm mm}$).
The on-source exposure times for the galaxies were estimated from their $150\,{\rm \mu m}$ continuum luminosities and total to $1.48\,{\rm h}$, $2.24\,{\rm h}$, $0.64\,{\rm h}$, and $0.41\,{\rm h}$.
For each target, the correlator was set up in dual polarisation to cover two spectral windows of $1.875\,{\rm GHz}$ bandwidth each at a resolution of $31.25\,{\rm MHz}$ ($\sim 23\,{\rm km\,s^{-1}}$) in each sideband and centered at $406.4\,{\rm GHz}$. 

The Common Astronomy Software Application (CASA) version 5.4.0 was used for data calibration and analysis. For the data calibration, we used the scripts released by the QA2 analyst ({\fontfamily{cmtt}\selectfont{ScriptForPI.py}}).
We then produced continuum maps using the CASA task {\fontfamily{cmtt}\selectfont{TCLEAN}} using multi-frequency synthesis ({\fontfamily{cmtt}\selectfont{MFS}}) mode with  {\fontfamily{cmtt}\selectfont{NATURAL}} weighting scheme to maximise their sensitivities.
During the {\fontfamily{cmtt}\selectfont{TCLEAN}} process, we deconvolved synthesised beam down to $3\,\sigma$, where $\sigma$ is the background RMS of the image without beam deconvolution (i.e. ``dirty image").
The resulting continuum sensitivities of the Band 8 maps are $34\,\rm{\mu Jy\,beam^{-1}}$, $40\,\rm{\mu Jy\,beam^{-1}}$, $57\,\rm{\mu Jy\,beam^{-1}}$, and $66\,\rm{\mu Jy\,beam^{-1}}$ for \textit{HZ4}, \textit{HZ6}, \textit{HZ9}, and \textit{HZ10} respectively.
All sources are significantly detected ($\rm{S/N}\simeq8-25$) as expected from our observing strategy (top panels, Figure \ref{fig:cutouts}).

\begin{figure*}
    \centering
    \includegraphics[width=0.95\textwidth]{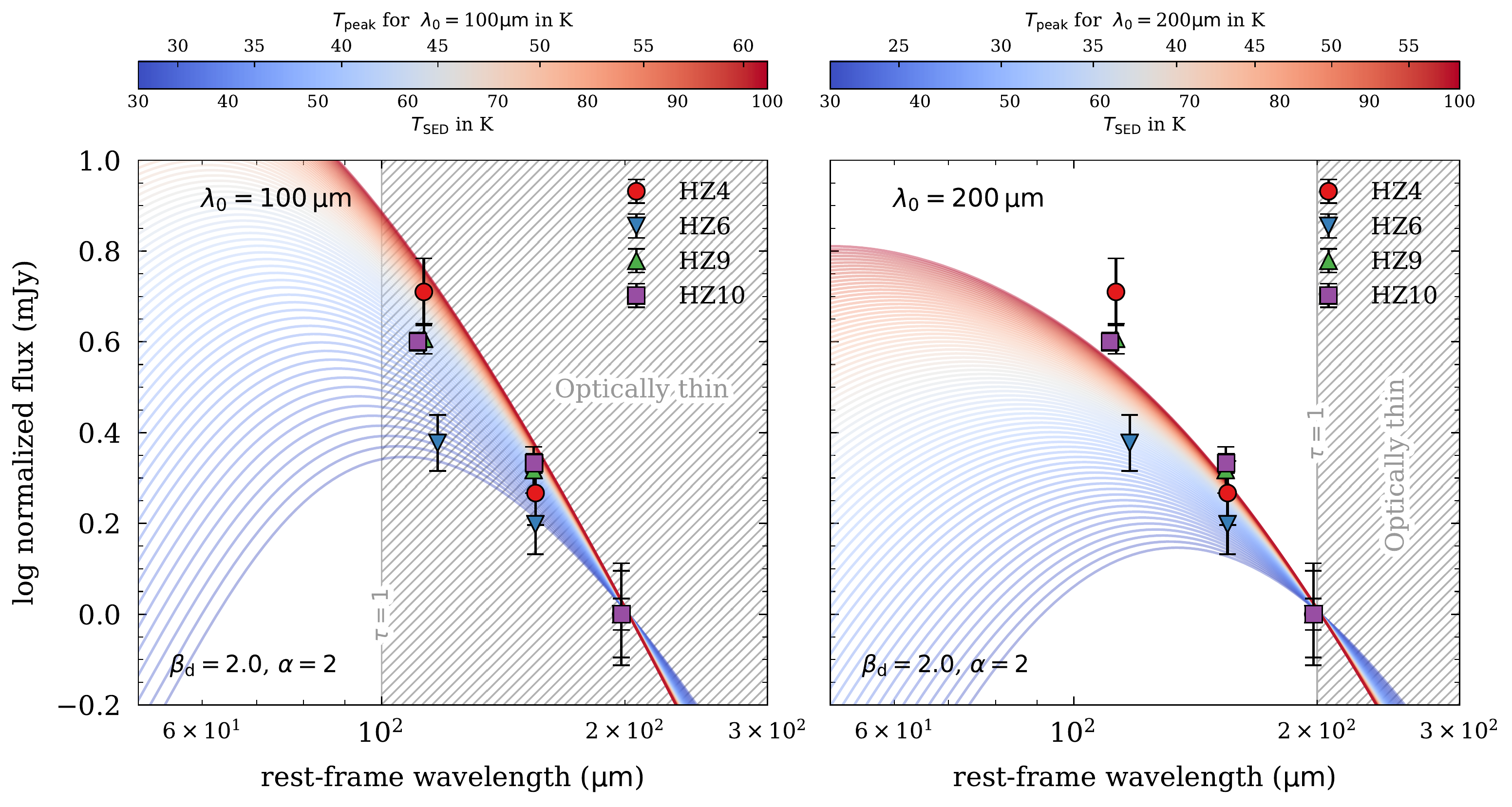}
    \caption{Infrared SED models assuming $\lambda_{0}=100\,{\rm \mu m}$ (left) and $\lambda_{0}=200\,{\rm \mu m}$ (right) and fixed $\alpha=2$ and $\beta_{\rm d}=2$ for different SED temperatures from $30\,{\rm K}$ (blue) to $100\,{\rm K}$ (red). The colour bar also indicated the dust peak temperature. The continuum fluxes of all galaxies except \textit{HZ6} can only be reproduced in the optical thin emission redward of $100\,{\rm \mu m}$ ($\lambda_{0}=100\,{\rm \mu m}$). This suggests optical thin dust emission. The observations are normalised to Band 6 (at $205\,{\rm \mu m}$).}
    \label{fig:opacity}
\end{figure*}

\subsection{Ancillary ALMA data}
\label{sec:ancillary}
To measure continuum at longer wavelengths, we complemented our Band 8 observations with ancillary data available in the ALMA archive. All four sources were observed by several observing projects both in Band 6 and in Band 7. We refer interested readers to the papers listed below for more detail.

The rest-frame $\sim200\,\rm{\mu m}$ (Band 6) observations are available from the ALMA project code 2015.1.00388.S \citep[PI: N. Lu,][]{LU18} and 2015.1.00928.S \citep[PI: Pavesi,][]{PAVESI19} and the rest-frame $\sim150\,\rm{\mu m}$ (Band 7) observations are available from the ALMA project code 2017.1.00428.L \citep[{\it ALPINE}; PI: O. Le F\`evre,][]{ALPINE_LEFEVRE19,ALPINE_BETHERMIN19,ALPINE_FAISST20} for {\it HZ4} and for {\it HZ6}, and 2012.1.00523.S \citep[PI: P. Capak,][]{CAPAK15} for {\it HZ9} and for {\it HZ10}.
After we obtained data from the ALMA archive, we calibrated all data using the scripts released by the QA2 analyst ({\fontfamily{cmtt}\selectfont{ScriptForPI.py}}). We use the appropriate versions of CASA as specified by the scripts.

To create continuum maps, we excluded channels closer than $\pm3\sigma$ width of the \nii~($205\,\rm{\mu m}$) or \Cii~($158\,\rm{\mu m}$) emission lines from the calibrated data.
The emission line frequencies and widths are taken from previous studies \citep{CAPAK15,PAVESI18,ALPINE_BETHERMIN19}. After masking emission lines, we create continuum maps following the same procedure as for the Band 8 data using CASA task {\fontfamily{cmtt}\selectfont{TCLEAN}} with {\fontfamily{cmtt}\selectfont{NATURAL}} weighting scheme (see \S \ref{sec:band8}).
The resulting sensitivities and synthesised beam resolutions are summarised in Table \ref{tab:observations}.
All maps show significant ($\rm{S/N} \simeq 5-30$) continuum detections, with spatial positions that are consistent across different frequencies given the beam uncertainties (bottom panels, Figure \ref{fig:cutouts}).

\subsection{Flux Calibration Errors}
\label{sec:calibrationerror}

Given the significant detections and high flux densities measured for some of our sources, flux calibration errors are potentially a significant contributor to the overall uncertainties. We thus estimated the variability of all our flux calibrators.
In particular, when flux calibrations are performed using secondary flux calibrators (i.e. quasars), we obtained the flux monitoring results from the ALMA calibrator source catalog\footnote{\url{https://almascience.eso.org/sc/}} both for Band 3 and for Band 7.
We then estimated expected flux densities and errors in each observed frequency.
The differences from the expected fluxes from each successive monitoring are used to estimate flux variabilities of observed frequencies.
In doing so, we accounted for the typical measurement uncertainties of the expected flux densities.
When flux calibrations are based on primary flux calibrators (i.e. solar systems objects), the flux calibrations are much less affected by the flux variabilities. Nevertheless, we applied conservative flux calibration errors of $5\,\%$ to take into account the potential modeling uncertainty of resolved flux calibrator observations.
We estimated flux calibration errors of $6\,\%$ to $\sim9\,\%$ for our observations (see Table \ref{tab:observations}).

\begin{table*}
\caption{Summary of quantities derived from the infrared SED. All values (except $\beta_{\rm d}$) are computed by marginalising over $\beta_{\rm d}$. The quoted errors include all uncertainties and are $1\sigma$.} \label{tab:fits}
\begin{tabular}{lccccccccc}
  \hline\hline
  ID & $\beta_{\rm d}$ & $T_{\rm SED}$ & $T_{\rm peak}$ & $\log{(L_{\rm IR})}^{a}$ & $\log{(L_{\rm FIR})}^{b}$ & $\log{(M_{\rm ISM})}$ & $f_{\rm ISM}$ & $\log({\rm SFR_{tot}})$ & $t_{\rm depl}$\\
  &  & $\left[\rm{K}\right]$ & $\left[\rm{K}\right]$ & $\left[\rm{L_{\odot}}\right]$ & $\left[\rm{L_{\odot}}\right]$ & $\left[\rm{M_{\odot}}\right]$ &  &  $\left[{\rm M_{\odot}\,yr^{-1}}\right]$ & $\left[\rm{Myrs}\right]$ \\ \hline
    {\it HZ4} & 2.01$^{+0.69}_{-0.57}$ & 57.3$^{+67.1}_{-16.6}$ & 42.4$^{+28.7}_{-8.2}$ & 11.91$^{+0.37}_{-0.91}$ & 11.83$^{+0.44}_{-0.23}$ & 9.92$^{+0.43}_{-0.50}$ & 0.67$^{+0.21}_{-0.31}$  & 1.89$^{+0.37}_{-0.91}$ & 97$^{+265}_{-70}$   \\
        {\it HZ6} & 1.60$^{+0.58}_{-0.57}$ & 40.8$^{+17.8}_{-7.2}$ & 33.9$^{+9.9}_{-5.9}$ & 11.73$^{+0.22}_{-0.34}$ & 11.52$^{+0.24}_{-0.31}$ & 10.55$^{+0.43}_{-0.45}$ & 0.73$^{+0.17}_{-0.26}$  & 1.72$^{+0.22}_{-0.34}$ & 755$^{+2041}_{-544}$ \\
        {\it HZ6}$^{\dagger}$ & 1.85$^{+0.69}_{-0.82}$ & 48.4$^{+30.2}_{-10.8}$ & 30.8$^{+15.4}_{-6.5}$ & 11.69$^{+0.26}_{-0.51}$ & 11.50$^{+0.29}_{-0.43}$ & 9.99$^{+0.78}_{-0.62}$ & 0.37$^{+0.49}_{-0.30}$  & 1.68$^{+0.26}_{-0.51}$ & 212$^{+173}_{-1094}$ \\
        {\it HZ9} & 2.01$^{+0.52}_{-0.70}$ & 49.4$^{+29.0}_{-10.7}$ & 38.9$^{+13.7}_{-5.8}$ & 12.14$^{+0.21}_{-0.45}$ & 11.94$^{+0.21}_{-0.33}$ & 10.39$^{+0.44}_{-0.35}$ & 0.81$^{+0.13}_{-0.19}$  & 2.13$^{+0.21}_{-0.45}$ & 210$^{+431}_{-141}$ \\
        {\it HZ10} & 2.15$^{+0.41}_{-0.54}$ & 46.2$^{+16.2}_{-8.5}$ & 37.4$^{+8.0}_{-4.9}$ & 12.49$^{+0.15}_{-0.25}$ & 12.28$^{+0.15}_{-0.21}$ & 10.72$^{+0.36}_{-0.26}$ & 0.70$^{+0.17}_{-0.16}$  & 2.48$^{+0.15}_{-0.25}$ & 191$^{+260}_{-107}$ \\ \hline
 \end{tabular}\\
 \begin{tablenotes}
            \item $\rm^{a}$ Total infrared luminosity computed in the range from $3\,{\rm \mu m}$ to $1100\,{\rm \mu m}$.
            \item $\rm^{b}$ Total far-IR luminosity computed in the range from $42.5\,{\rm \mu m}$ to $122.5\,{\rm \mu m}$.
            \item $\rm^{\dagger}$ Measurement with $\lambda_{0} = 200\,{\rm \mu m}$, see text for more details.
        \end{tablenotes}
\end{table*}


\section{Measurements}
\label{sec:measurements}

\subsection{Continuum Flux Measurements}
\label{sec:flux}

After we confirmed individual detections in all images, we performed continuum flux density measurements in the visibility domain.
While imaged maps are useful to examine the achieved sensitivities and to validate source detections, map reconstructions depend on observational and imaging parameters such as the resolution and parameters used during the deconvolution processes. The visibility domain is less affected by these parameters.

We performed visibility-based flux measurements
using the task {\fontfamily{cmtt}\selectfont{UV\_FIT}} from the software package GILDAS\footnote{GILDAS is an interferometry data reduction and analysis software developed by Institut de Radioastronomie Millim\'{e}trique (IRAM) and is available from \url{http://www.iram.fr/IRAMFR/GILDAS/}.
To convert ALMA measurement sets to GILDAS/MAPPING uv-table, we followed \url{https://www.iram.fr/IRAMFR/ARC/documents/filler/casa-gildas.pdf}.} after creating continuum visibilities by masking emission lines, if present, following the same procedure as in \S\ref{sec:ancillary}.
We used a single 2D Gaussian for visibility fitting, keeping source positions, source sizes, and integrated flux densities as free parameters.
The resulting measurements for all of our sources are listed in the Table~\ref{tab:measurements}.

\subsection{Infrared SED Fits and Dust Temperature}
\label{sec:analysis}

We use the Markov Chain Monte Carlo (MCMC) method provided by the Python package PyMC3\footnote{\url{https://docs.pymc.io/}} to fit the infrared SEDs of our four galaxies including all the three ALMA continuum measurements described above (Table~\ref{tab:measurements}). The SED is parameterised as the sum of a single modified black body and a mid-infrared power-law as described in \cite{CASEY12} \citep[see also][]{BLAIN03},

\begin{equation}\label{eq:firsed1}
S(\lambda) = N_{\rm bb}\,f(\lambda;\beta_{\rm d},T_{\rm SED}) + N_{\rm pl}\, \lambda^\alpha\, e^{-(\lambda/\lambda_{\rm c})^2}
\end{equation}

with 

\begin{equation}\label{eq:firsed2}
    N_{\rm pl} \equiv N_{\rm bb} \, f(\lambda_{\rm c};\beta_{\rm d},T_{\rm SED})
\end{equation}

and

\begin{equation}\label{eq:firsed3}
    f(\lambda;\beta_{\rm d},T_{\rm SED}) \equiv \frac{ \left( 1 - e^{-( \lambda_0 / \lambda)^{\beta_{\rm d}}} \right) \left( \frac{c}{\lambda} \right)^3 }{e^{ (h\,c)/(\lambda\, k\, T_{\rm SED}) } - 1}.
\end{equation}

\begin{figure}
    \centering
    \includegraphics[width=0.46\textwidth]{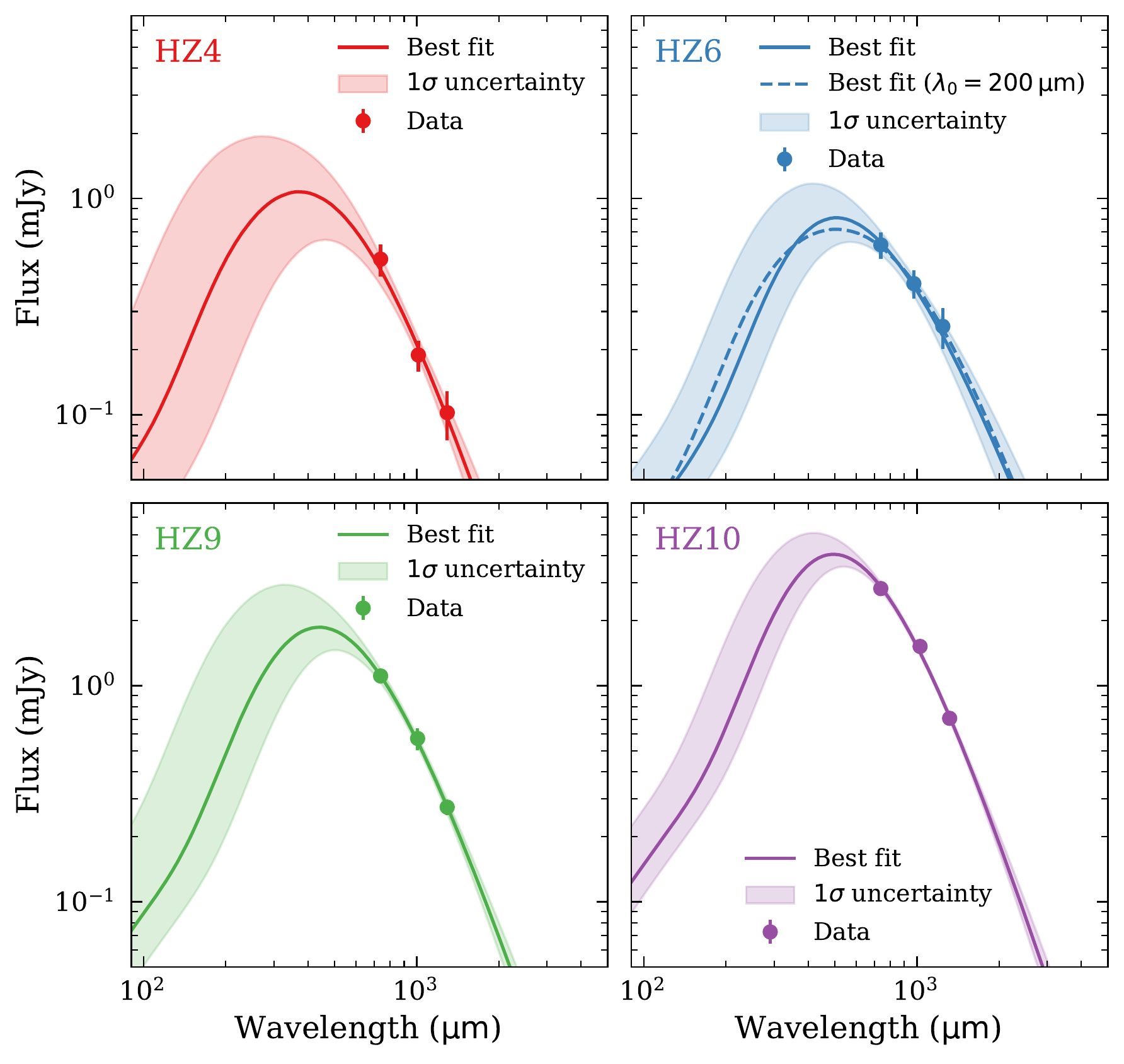}
    \caption{Best-fit infrared SEDs for the four galaxies derived from Equations~\ref{eq:firsed1} to~\ref{eq:firsed3} ($1\sigma$ uncertainty indicated by coloured band). The wavelength is given in observer frame. The dashed line shows the case with $\lambda_0=200\,{\rm \mu m}$ for \textit{HZ6}. The new Band 8 continuum measurements ($110\,{\rm \mu m}$ rest-frame, $740\,{\rm \mu m}$ observed) probe better the peak of the infrared SED, which allows us to put the first constraints on the dust temperature and total infrared luminosity at these redshifts. Note that the scale of the $y-$axis is the same in all panels to show the differences in total flux.}
    \label{fig:SED}
\end{figure}

In addition, the power-law turnover wavelength $\lambda_{\rm c}$ is dependent on $\alpha$ and $T_{\rm SED}$ \citep[see][]{CASEY12}.
Free parameters are $N_{\rm bb}$ (normalisation), $\alpha$ (slope of the mid-infrared power-law), $\beta_{\rm d}$ (emissivity index), $T_{\rm SED}$ (SED dust temperature), and $\lambda_{0}$ (wavelength where the optical depth is unity). Since our data do not constrain the SED blueward of rest-frame $\sim110\,{\rm \mu m}$ (Band 8), we fix the mid-IR power-law slope to $\alpha=2.0$ as suggested by the measurements in \citet[][]{CASEY12}.

The SED dust temperature (defined by Equation~\ref{eq:firsed1}) should not be confused with the peak dust temperature, which is proportional to the inverse wavelength at the peak of the infrared emission via Wien's displacement law,

\begin{equation}\label{eq:tpeak}
    T_{\rm peak}\,{\rm [K]} = \frac{ 2.898\times10^3\,{\rm [\mu m\,K]}}{ \lambda_{\rm peak}\,{\rm [\mu m]}}.
\end{equation}

The SED and peak temperature can be considerably different as shown in \citet[][]{CASEY12}. Note that both are a measure of the \textit{light-weighted} dust temperature. This is in contrast to the cold dust emitted at $25\,{\rm K}$ in the Rayleigh-Jeans tail of the far-IR spectrum ($\gtrsim~250\,{\rm \mu m}$ rest-frame). This \textit{mass-weighted} temperature is expected to be largely independent of redshift and other galaxy properties \citep[see, e.g.,][]{SCOVILLE16,LIANG19}.

\begin{figure*}
    \centering
    \includegraphics[width=0.99\textwidth]{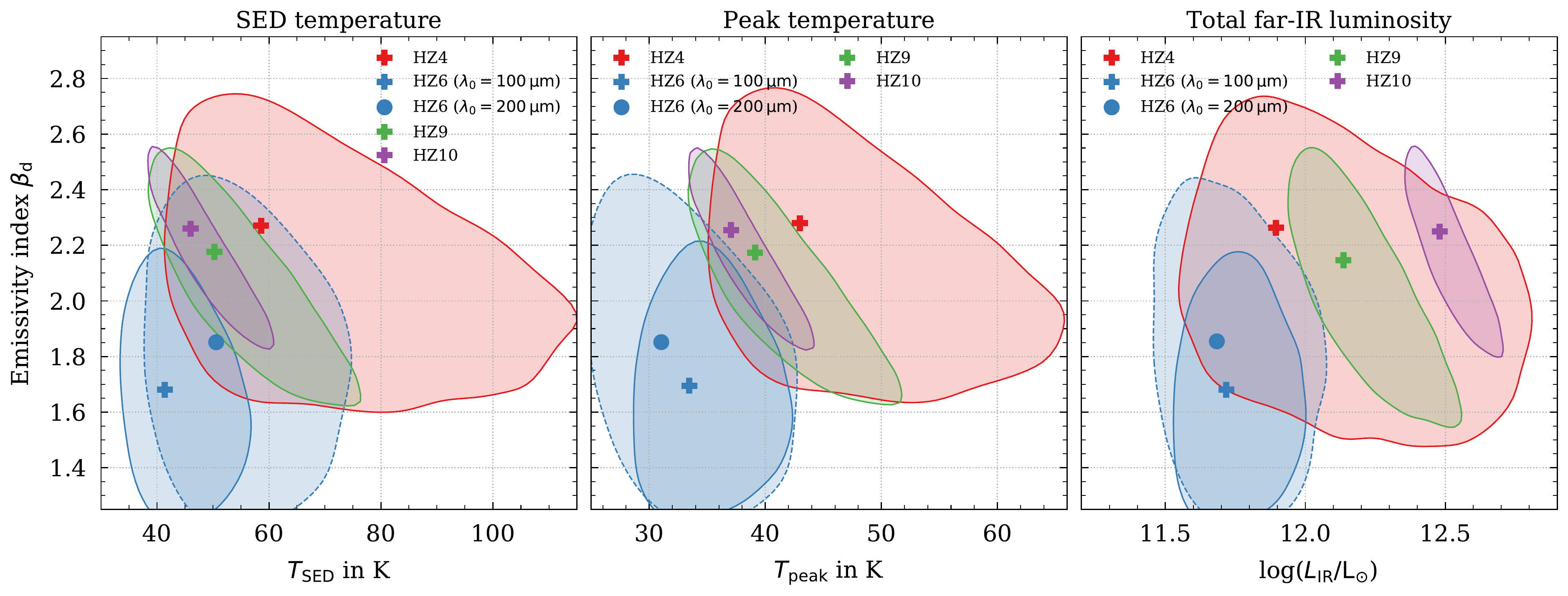}
    \caption{MCMC-derived $1\sigma$ contours of the SED (left) and peak (middle) dust temperature as well as the total ($3-1100\,{\rm \mu m}$) infrared luminosity (right) as a function of the emissivity index $\beta_{\rm d}$. The dashed contour shows the results for \textit{HZ6} in the optically thick case ($\lambda_0=200\,{\rm \mu m}$). Band 8 probes the wavelength close to the far-IR peak for our galaxies, hence allows us to put first constraints on dust temperatures at these redshifts.}
    \label{fig:temperaturetraces}
\end{figure*}

A largely unknown fitting parameter is $\lambda_{0}$, the wavelength at which the optical depth $\tau$ equals unity (i.e., optically thick at bluer wavelengths). Based on observational studies at lower redshift, it is generally assumed that $\lambda_0 \sim 200\,{\rm \mu m}$ \citep[e.g.,][]{BLAIN03,CONLEY11,RANGWALA11,CASEY12,RIECHERS13}. 
However, as shown in Figure~\ref{fig:opacity}, our new Band 8 observations cannot be fit with $\lambda_0=200\,{\rm \mu m}$ for three out of four galaxies. Specifically, the two panels show rest-frame modified black body models (Equations~\ref{eq:firsed1} to~\ref{eq:firsed3} with fixed $\alpha=2$ and $\beta_{\rm d}=2$) for $\lambda_{0}=100\,{\rm \mu m}$ (left) and $\lambda_{0}=200\,{\rm \mu m}$ (right) for a range of SED temperatures (coloured from blue to red). The observed fluxes of our galaxies normalised to Band 6 (at $205\,{\rm \mu m}$) are shown by symbols.
Clearly, our Band 8 observations (at rest-frame $110\,{\rm \mu m}$) cannot be explained with $\lambda_{0}=200\,{\rm \mu m}$ at any reasonable temperature for all of our galaxies except \textit{HZ6}. The emission at $\sim100\,\mu m$ is therefore likely optically thin and we therefore assume $\lambda_{0}=100\,{\rm \mu m}$ in the following. This is consistent with theoretical models for low-opacity dust \citep{DRAINE06,SCOVILLE76}. Different values of $\alpha$ and $\beta_{\rm d}$ in a reasonable range do not change this conclusion.

The observations of \textit{HZ6} can be reconciled with optically thick emission up to rest-frame $200\,{\rm \mu m}$. As found in \citet[][]{CAPAK11}, \textit{HZ6} is part of a protocluster at $z=5.3$. Specifically, \textit{HZ6} consists of three components separated by $\Delta v < 50\,{\rm km\,s^{-1}}$ in radial velocity and $< 3\,{\rm kpc}$ in projected distance (Figure~\ref{fig:cutouts}). The components are likely gravitationally interacting and a past close passage is suggested by the diffuse rest-frame UV emission and a `crossing time' of $\sim 50\,{\rm Myrs}$. The latter is estimated using $t_{\rm cross} \simeq (G\,\bar{\rho})^{-\frac{1}{2}}$, where $\bar{\rho}$ is the average mass density and $G$ the gravitational constant, with values based on observations ($r=3\,{\rm kpc}$ and total enclosed mass of $10^{10}\,{\rm M_{\odot}}$ for a single component). This setup could cause a more optically thick medium by, e.g., the compression of gas and/or the formation of dust. With the current data, it is not possible to make further conclusions and we therefore show in the following derivations using both values of $\lambda_0$ for \textit{HZ6}.

For the MCMC fit to the infrared SEDs of our galaxies, we adopt a flat prior for the dust temperature, and a Gaussian prior for $\beta_{\rm d}$ with a $\sigma(\beta_{\rm d})=0.5$ centered on $1.8$ \citep[see, e.g.,][]{HILDEBRAND83}. The normalisation is also sampled with a Gaussian prior in linear space around an initial guess derived by the normalisation in Band 7. We found that fitting in linear space is more appropriate given the errors of the data.
To perform the fitting, we use the \textit{No-U-Turn Sampler} \citep[NUTS,][]{HOFFMAN11}, which is an extension to the Hamilton Monte Carlo algorithm \citep[][]{NEAL12} and is less sensitive to tuning. We draw $18\,000$ samples in total with a target acceptance of $0.99$, which we found to provide the best performance.

Figure~\ref{fig:SED} shows the best-fit infrared SEDs together with the $1\sigma$ uncertainties for each of our galaxies. Thanks to our Band 8 data at rest-frame wavelengths of $110\,{\rm \mu m}$, we can put more stringent constraints on the location of the peak of the infrared SED. The galaxies \textit{HZ4} and \textit{HZ6} are fainter, resulting in larger uncertainties of the fit. While the mid-IR blueward of the peak is poorly constrained, the RJ tail (at $>1000\,{\rm \mu m}$ observed frame) can be robustly extrapolated based on our data.

Figure~\ref{fig:temperaturetraces} shows the derived SED (left) and peak (middle) dust temperature as well as total infrared luminosity (right) contours ($1\sigma$) as a function of the emissivity index $\beta_{\rm d}$ for our four galaxies.
We find emissivity indices between $1.6$ and $2.4$ for all galaxies, with a median of $2.0$, which is consistent with measurements at lower redshifts \citep[e.g.,][]{CASEY12,CONLEY11}. The dust SED temperatures range between $40-60\,{\rm K}$ with a median at $48\,{\rm K}$.
For the dust peak temperatures, we find a range of $30-43\,{\rm K}$ with a median of $38\,{\rm K}$.
The total infrared luminosities ($\LIR$) are derived by integrating the best-fit SED between $3-1100\,{\rm \mu m}$ and range between $5-30\times10^{11}\,{\rm L_{\odot}}$.
We note that the peak temperature and the total infrared luminosity is insensitive to the assumed $\lambda_0$.
We also quote far-IR luminosities ($\LFIR$) measured by integrating the flux between $42.5\,{\rm \mu m}$ and $122.5\,{\rm \mu m}$ for easier comparison with the literature.
All measurements are summarised in Table~\ref{tab:fits}.

\begin{figure*}
    \centering
    \includegraphics[width=0.71\textwidth]{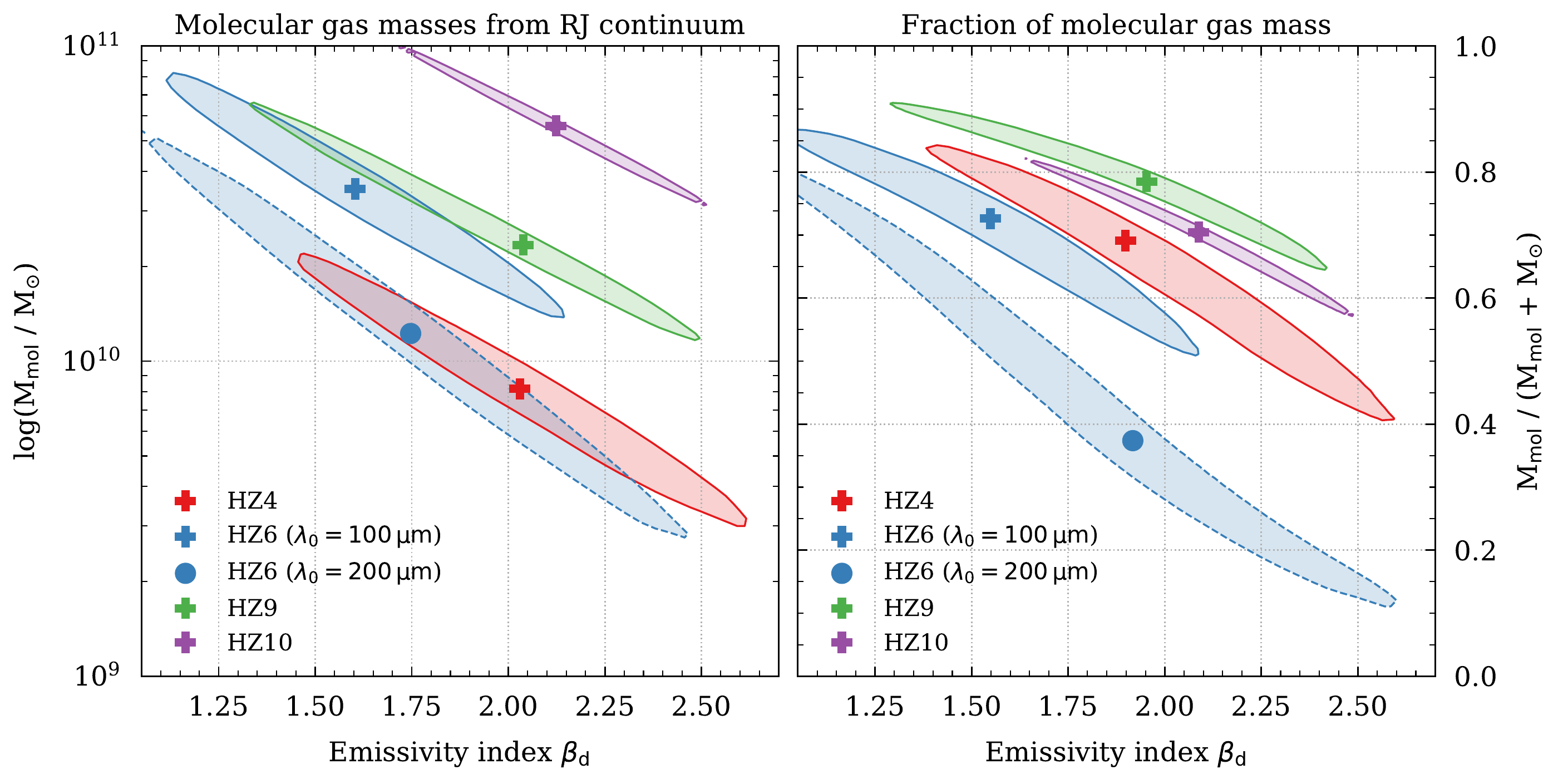}
    \includegraphics[width=0.27\textwidth]{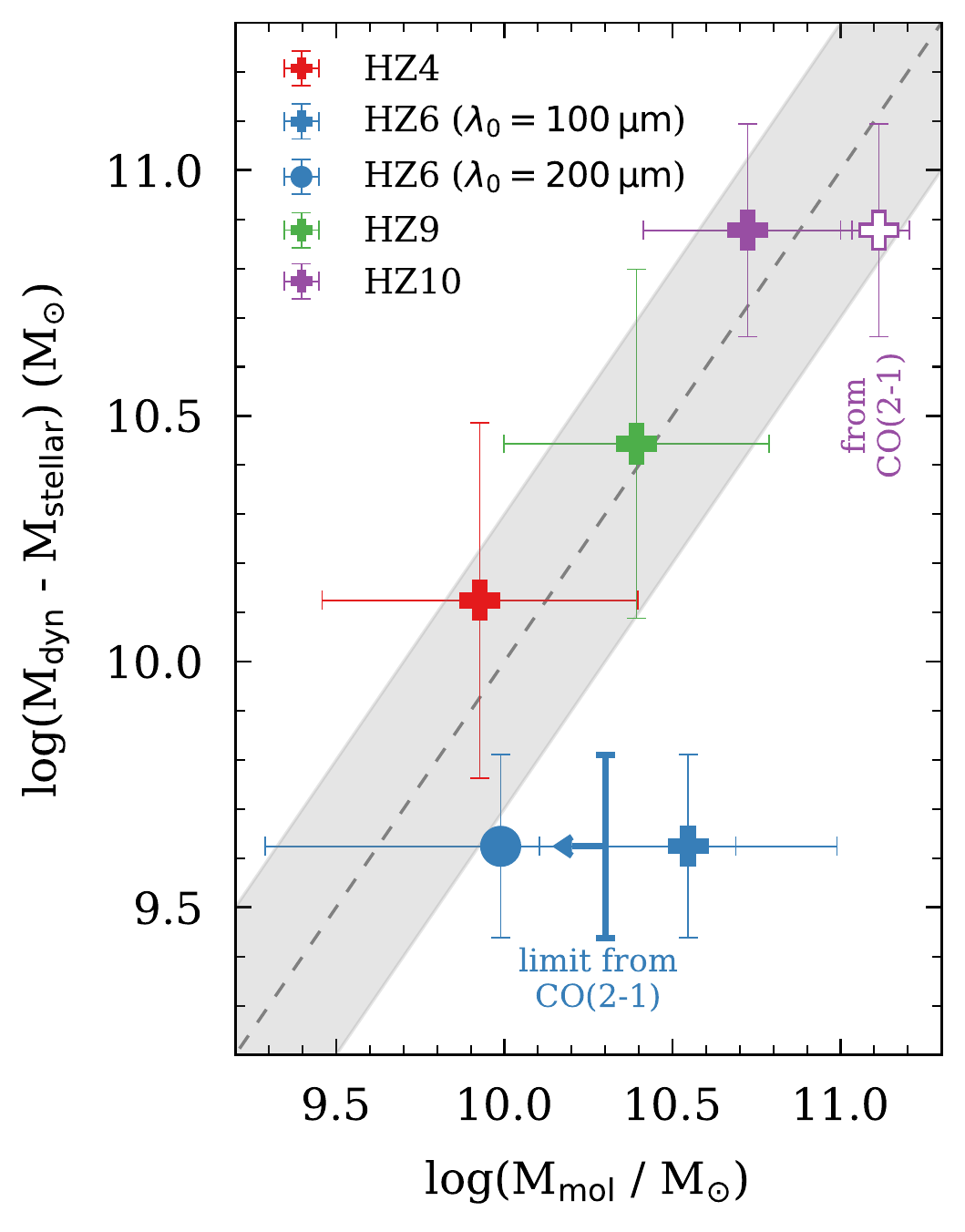}
    \caption{MCMC-derived $1\sigma$ contours of the molecular gas masses (left) and fractions (middle) as a function of the emissivity index $\beta_{\rm d}$. The molecular gas masses have been estimated using the RJ dust continuum at rest-frame $850\,{\rm \mu m}$ using the method by \citet{SCOVILLE16}. The three ALMA continuum frequencies put the first robust constraints on the emissivity index (far-IR slope) at these high redshifts, which is critical to derive molecular gas masses. The right panel shows the comparison of our dust-continuum derived molecular gas masses and the difference between dynamical and stellar mass (expected to be equal to the total gas mass modulo dark matter). Note that \textit{HZ6} (blue) has a complicated velocity structure due to its major merger nature, which makes the measurement of its dynamical mass significantly uncertain. The molecular gas mass measurements obtained from the CO($2-1$) transition \citep{PAVESI19} are also shown for comparison. They largely agree with our measurements.}
    \label{fig:gasmass}
\end{figure*}

\subsection{Molecular Gas Masses from the RJ dust continuum}\label{sec:Mism}

The measurement of molecular gas masses of galaxies is crucial to understand the star formation processes determining their growth and evolution. Low-$J$ transitions of the CO molecule are used regularly at $z<2$ \citep[e.g.,][]{TACCONI10,GENZEL15,FREUNDLICH19}, but this is not feasible for large samples of normal galaxies at higher redshifts due to the large amount of necessary telescope time. Currently only very few observations of CO in normal galaxies at $z>5$ exist \citep[][]{DODORICO18,PAVESI19}. Alternatively, the far-IR \Cii~emission line can be used as tracer of molecular gas \citep{DELOOZE14,ZANELLA18,ALPINE_DESSAUGES20}, however, there are considerable uncertainties in its use due to the unknown origin of C$^+$ emission.

Alternatively, molecular gas masses can be measured using the dust continuum emission emitted at rest-frame $850\,{\rm \mu m}$ in the RJ tail of the far-IR spectrum \citep{SCOVILLE14,SCOVILLE16,SCOVILLE17,HUGHES17,KAASINEN19,ALPINE_DESSAUGES20}.
For current samples of main-sequence galaxies at high redshifts, the far-IR slope (defined by the emissivity index $\beta_{\rm d}$) cannot be constrained directly due to the lack of observations. Significant assumptions have therefore to be made to quantify the rest-frame $850\,{\rm \mu m}$ continuum. With our 3-band data sampling the SED redward of the far-IR peak, we can constrain the far-IR slope for the first time at these redshifts directly.
The molecular gas masses are then derived using the observed flux at rest-frame $205\,{\rm \mu m}$ (Band 6, $S_{\rm \nu_{obs}}$ in mJy\footnote{Note that the dust is likely optically thin at this wavelength, c.f. Figure~\ref{fig:opacity}.}) that is extrapolated to $805\,{\rm \mu m}$ using the full probability distribution of $\beta_{\rm d}$ from our MCMC fit and equation (16) in \citet[][]{SCOVILLE16} with similar assumptions,

\begin{align*}\label{eq:mass}
M_{\rm mol} &= 1.78\,S_{\rm \nu_{obs}}\,(1+z)^{-(3+\beta_{\rm d})}\,\left( \frac{\nu_{\rm 850\mu m}}{\nu_{\rm obs}} \right)^{2+\beta_{\rm d}}\,D_{\rm L}^2(z)\\
&\times \left(\frac{6.7 \times 10^{19} }{\alpha_{850}} \right)\,\left(\frac{\Gamma_{\rm RJ,z=0}}{\Gamma_{\rm RJ}}\right) \times 10^{10}\,{\rm M_{\odot}}.
\end{align*}

In this case, $D_{\rm L}(z)$ is the luminosity distance at redshift $z$ in Gpc and we assume $\alpha_{850} = 6.7 \times 10^{19}\,{\rm erg\,s^{-1}\,Hz^{-1}\,M_{\odot}^{-1}}$, which is the average measured for galaxies at $z<3$ \citep{SCOVILLE16}. $\Gamma_{\rm RJ}(\nu_{\rm obs},T_{d},z)$ is the correction for departure in the rest-frame of the Planck function from Rayleigh-Jeans and depends on the \textit{mass-weighted} dust temperature (different from $T_{\rm SED}$ or $T_{\rm peak}$, which are \textit{luminosity-weighted} temperatures). For the latter, we adopt $25\,{\rm K}$, but assuming higher temperatures such as $35\,{\rm K}$ lowers the inferred molecular masses by less than $10\%$.

The left panel of Figure~\ref{fig:gasmass} shows the derived $1\sigma$ contours of the molecular masses for our galaxies from our MCMC fit. The masses range between $0.3-8.0\times 10^{10}\,{\rm M_{\odot}}$. The middle panel shows the molecular gas fractions ($f_{\rm gas} = M_{\rm mol} / (M_{\rm mol}+M_{\rm stellar})$, using stellar masses from \citet[][]{CAPAK15}), which range between $30\%$ and $80\%$. Statistically, this is consistent with the \textit{ALPINE} sample at $z=5.5$ \citep{ALPINE_DESSAUGES20}. This is expected as our galaxies are consistent with the average masses and SFRs of the \textit{ALPINE} sample \citep[see][]{ALPINE_FAISST20}.
The right panel of Figure~\ref{fig:gasmass} compares the dust continuum gas masses with the difference between dynamical and stellar masses. This difference should yield total gas masses modulo the contribution of dark matter, which is expected to be on the order of $10-20\%$ or less at the radii probed here \citep{BARNABE12}.
The dynamical masses are derived inside a half-light radius from the \Cii~emission line velocity profile \citep{PAVESI19}. 
Generally, we find an agreement within a factor of two ($<1\sigma$) between gas masses derived from dust continuum and dynamical masses. However, assuming $\lambda_0=100\,{\rm \mu m}$ for the fit of \textit{HZ6} results in a $2\sigma$ discrepancy. As noted earlier, \textit{HZ6} is a three-component major merger system with significant gravitational interaction. The complex velocity structure likely causes large uncertainties in its dynamical mass estimate. In addition, the $850\,{\rm \mu m}$-continuum derived gas mass encompasses the whole extended system, while the dynamical mass captures only a fraction of the gas. Both can explain its larger offset from the 1-to-1 line compared to the other galaxies. On the other hand, if the dynamical mass is reliable, this indicates once more that the emission in \textit{HZ6} could be optically think up to $200\,{\rm \mu m}$ (as $\lambda_0=200\,{\rm \mu m}$ results in discrepancy).

\citet[][]{PAVESI19} report a gas mass estimate from CO($2-1$) emission for \textit{HZ6}\footnote{This galaxy is named \textit{LBG-1} in their paper, see also \citet{RIECHERS14}.} and \textit{HZ10}, assuming a Milky Way-like CO to molecular gas conversion factor ($\alpha_{\rm CO} = 4.5\,{\rm M_{\odot} / (K\,km\,s^{-1}\,pc^2 )}$) and brightness temperature ratio $R_{21}=1$. They find $1.3 \times 10^{11}\,{\rm M_{\odot}}$ for \textit{HZ10} and a limit $<2 \times 10^{10}\,{\rm M_{\odot}}$ for \textit{HZ6} (see Figure~\ref{fig:gasmass}). The CO-gas mass estimate of \textit{HZ10} is a factor $2.5$ larger ($\sim1\sigma$ discrepancy) than what we measure from dust continuum and dynamical masses. The upper limit in CO-derived gas mass for \textit{HZ6} is consistent with our measurement if assuming $\lambda_0=200\,{\rm \mu m}$ (but not if optically thin dust at $100\,{\rm \mu m}$).
The discrepancy of the measurements for \textit{HZ10} are not significant, given the large measurement errors as well as uncertainties in $\alpha_{\rm CO}$ and the brightness temperature ratio. However, large $\alpha_{\rm CO}$ values above $20$ that are expected for metal-poor environments such as in the Small Magellanic Cloud \citep[see][]{LEROY11} can be excluded for \textit{HZ10}. This is in agreement with earlier studies that suggest that \textit{HZ10} is fairly metal enriched, even close to solar metallicity, based on its strong rest-frame UV absorption lines \citep[e.g.,][]{FAISST17b,PAVESI19}.
Clearly, larger samples or more precise measurements have to be obtained in order to draw final conclusions.

From the total infrared luminosity we can derive the total SFRs using the \citet{KENNICUTT98} relation and from that the gas depletion time via $t_{\rm depl} = \rm M_{\rm mol}$/SFR$_{\rm IR}$. For \textit{HZ4}, \textit{HZ9}, and \textit{HZ10} find values between $100-220\,{\rm Myrs}$. This is in agreement with the trend of decreasing depletion time with higher redshifts \citep[e.g.,][]{SCOVILLE16}. For \textit{HZ6}, the depletion time depends on the assumed $\lambda_0$. For $\lambda_0$ equal to $100\,{\rm \mu m}$ and $200\,{\rm \mu m}$ we derive a depletion time of $755\,{\rm Myrs}$ and $212\,{\rm Myrs}$, respectively.

\begin{figure*}
    \centering
    \includegraphics[width=0.95\textwidth]{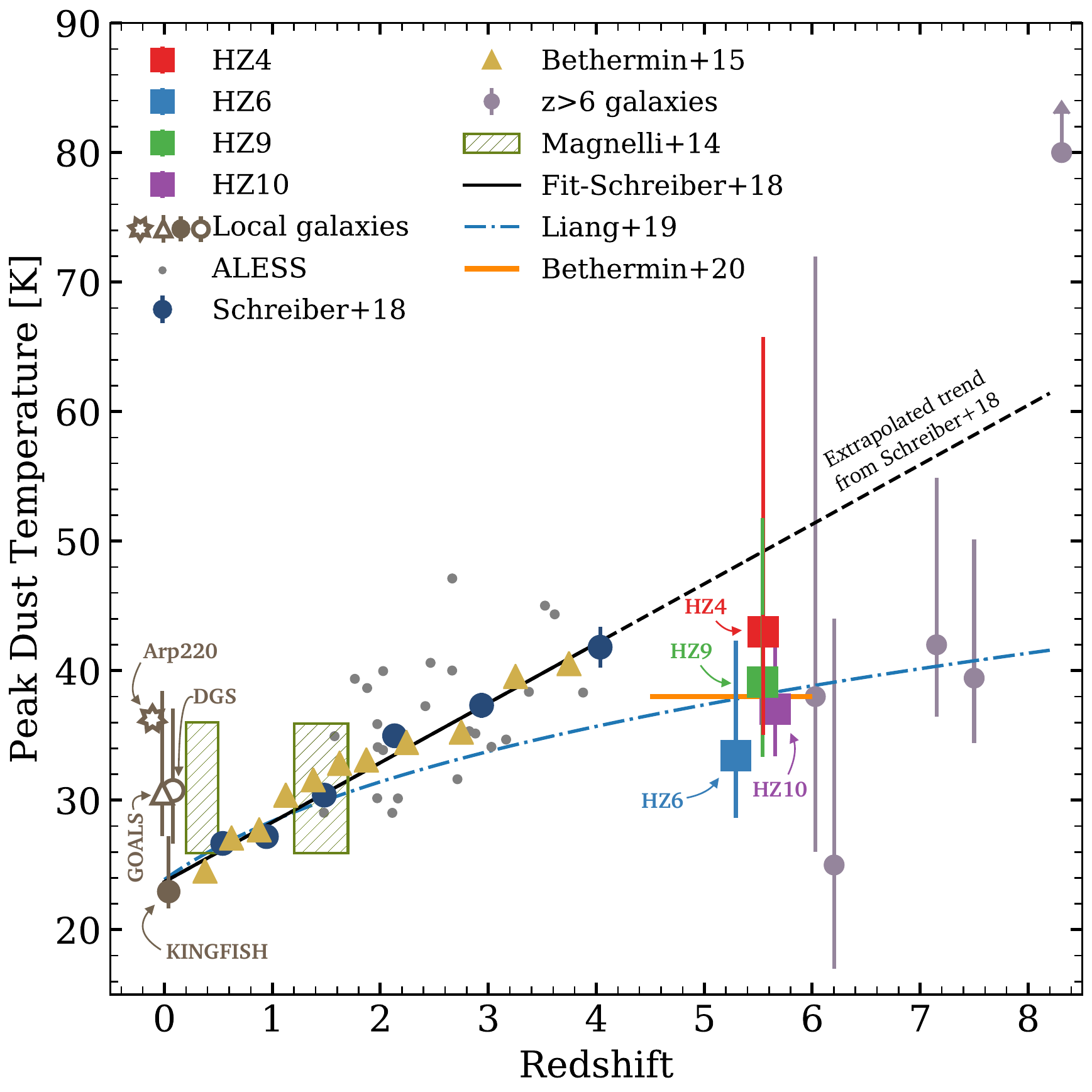}
    \caption{Peak dust temperature ($T_{\rm peak}$) evolution with redshift. The large coloured squares show our galaxies at $z\sim5.5$. We also show galaxy samples at $z=0$ from the KINGFISH (dark gray circle), DGS (dark gray open circle), and GOALS (dark gray open triangle) samples (shifted by $0.02$ in redshift for clarity), and galaxies at $z=0.2-4.0$ (matched to our luminosity range) from \textit{ALESS} \citep[gray small circles,][]{SMAIL14,SCHREIBER18}, \citet[yellow triangles,][]{BETHERMIN15}, and \citet[hatched rectangles,][]{MAGNELLI14}, as well as $z>6$ \citep[light purple circles,][]{KNUDSEN16,HASHIMOTO19,BAKX20}. The data for Arp220 is taken from \citet[][]{RANGWALA11}. For consistency, we re-measured the $T_{\rm peak}$ of the local galaxies as well as $z>6$ galaxies with our method (see Section~\ref{sec:measurements}). The latter include CMB correction.
    The fit to the \citet[][]{SCHREIBER18} data is shown in black (dashed when extrapolation), together with the expectation from hydrodynamic simulations \citep[blue dot-dashed,][]{LIANG19,MA19}, and the temperature derived from an average template for \textit{ALPINE} galaxies \citep[orange line,][]{ALPINE_BETHERMIN19}.
    Galaxies at $z>4$ have warmer peak temperatures compared to average local galaxies, however, are $5-10\,{\rm K}$ cooler than what would be predicted from the trend found at $z<4$. This indicates a flattening of the $T_{\rm peak}-z$ relation at $z>4$, which could be due to a lower dust abundance or opacity in high redshift galaxies.}
    \label{fig:tempevolution}
\end{figure*}

\section{Discussion}\label{sec:discussion}

\subsection{Rising Temperature towards High Redshifts?}

Figure~\ref{fig:tempevolution} puts our measurements at $z\sim5.5$ into context with measurements from the literature at $z<4$ and $z>6$.
At $z=0$, we show peak temperature measurements derived in \citet[][using Equations~\ref{eq:firsed1} to~\ref{eq:firsed3}]{FAISST17b} for the KINGFISH sample \citep{SKIBBA11}, the \textit{Dwarf Galaxy Sample} \citep[DGS,][]{MADDEN13}, and the GOALS sample \citep{KENNICUTT11}. The data for Arp220 is taken from \citet[][]{RANGWALA11}.
At $z=0.2-4.0$ from the \textit{ALMA LABOCA ECDFS Sub-mm Survey} \citep[ALESS,][]{SMAIL14,SCHREIBER18}, the sample from \citet[][]{BETHERMIN15}, and galaxies at from \citet[][]{MAGNELLI14}. For the latter, we select a similar range in total infrared luminosity as our sample ($\log(\LIR/{\rm L_{\odot}}) \sim 10.5-12.5$). We also note that the stellar mass range is similar to our sample.
At $z>6$ we show galaxies from \citet[][lower limit]{BAKX20}, \citet[][]{KNUDSEN16}, and \citet[][]{HASHIMOTO19}. For the latter two, we have re-measured the peak temperature with our method. 
The trend derived from hydrodynamic simulations \citep[][]{LIANG19,MA19} is shown as dot-dashed line.

Our measurements at $z\sim 5.5$ show elevated dust temperatures compared to average local galaxies such as from the KINGFISH sample. Taking the uncertainties into account, we can exclude peak temperatures of less than $30\,{\rm K}$. The median peak temperature measured for our galaxies is at $38\pm5\,{\rm K}$, with a low-probability tail towards higher temperatures (Figure~\ref{fig:temperaturetraces}).
However, although the temperatures at $z\sim5.5$ are higher compared to average local galaxies, we find that our values are about $10\,{\rm K}$ below what would be predicted from an extrapolation of observational data at $z<4$. Particularly, the \citet[][]{SCHREIBER18} infrared SED template at $z=4$ suggests a peak temperature of $42\,{\rm K}$. If extrapolated to $z=5.5$, this results in $\sim46\,{\rm K}$, which is $\sim5-10\,{\rm K}$ higher than what we measure from our data (at similar infrared luminosity).
On the other hand, our measurements are consistent with the empirically derived infrared SED of $4 < z < 6$ galaxies from \citet[][]{ALPINE_BETHERMIN19}, who find an average peak temperature of $38\,{\rm K}$. Furthermore, we find similar peak temperatures as reported at $z>6$ by the various studies.

Summarising, the peak temperatures of high-$z$ galaxies are warmer compared to average local galaxies, which has to be taken into account when parameterising the infrared SEDs of these galaxies. However, our observational data suggest that the peak temperature (i.e., the wavelength at peak emission of the infrared SED) does not evolve anymore strongly beyond redshifts $z=4$ for a fixed total infrared luminosity.
This behaviour is reproduced in hydrodynamic simulations \citep[e.g.,][]{LIANG19,MA19}, which show a flattening of the temperature evolution with redshift for galaxies selected with $\LIR > 10^{11}\,{\rm L_{\odot}}$ (Figure~\ref{fig:tempevolution}).

\subsection{Explaining the Observed $T_{\rm peak}(z)$ Evolution\\with an Analytical Model of a Spherical Dust Cloud}\label{sec:opacity}

In the previous section we have constrained the evolution of the peak temperature with redshift. Our unique observations at $z\sim5.5$ and literature data at lower redshifts suggest that the temperature rises up to $z\sim4$ and then tends to flatten off.
At this point, we note that the increase in SED or peak temperature with redshift is merely a statement on a shift in the wavelengths at which the infrared SED peaks, i.e., its \textit{shape}. This shift can be due to several physical reasons, including changes in the UV luminosity of a central source (i.e., the young stars), the dust mass density or the opacity of the dust. 
Such dependencies have been seen observationally in the local galaxy samples (Figure~\ref{fig:tempevolution}). For example, the KINGFISH sample (consisting of mostly solar metallicity and infrared fainter ($10^{10}-10^{11}\,{\rm L_{\odot}}$) galaxies) shows peak temperatures between $20\,{\rm K}$ and $30\,{\rm K}$. On the other hand, the IR luminous ($10^{11}-10^{12}\,{\rm L_{\odot}}$) GOALS sample (also close to solar metallicity) shows higher average peak temperatures ($25-40\,{\rm K}$). The DGS sample shows similarly high peak temperatures as the GOALS sample at less than a tenth solar metallicity and $\LIR<10^{11}\,{\rm L_{\odot}}$. This suggests that $T_{\rm peak}$ correlates negative with metallicity and positive with $\LIR$ \citep[see figure 4 in][]{FAISST17b}. 
These physical relations can make selection effects (such as a survey limit in total infrared luminosity) mimic a temperature increase at $z<4$. However, as shown in table 1 of \citet{SCHREIBER18}, even at a fixed total infrared luminosity, the trend of increasing peak temperatures from $25\,{\rm K}$ to $40\,{\rm K}$ at $z=0.3-4.0$ persists.

In the following, we use a simple analytical model to investigate how the output of UV photons from a dust-enshrouded source (specifically the ratio between UV luminosity and dust mass) and the density of dust (i.e., the dust opacity) affect the shape of the infrared SED and with it the emergent peak temperature measured by an external observer.

We model the emitted heat (and hence peak temperature) from a dust cloud around stars using a model based on \citet[][]{SCOVILLE76} that will be described in a forthcoming work (Scoville et al. in prep.). The model assumes a central source of UV light enshrouded in a dust cloud with spherical symmetry and constant density, and calculates the heating in concentric shells of dust mass. Secondary heating (from re-emitted light) is included, as well as the increased background temperature by CMB heating at high redshift. The latter, however, does not affect the dust temperature below $z=6$ significantly.
In the following, we consider models for different intrinsic (i.e., obscured plus unobscured) UV luminosities and dust cloud radii. For the UV luminosity we choose $10^{11}$ and $10^{12}\,{\rm L_{\odot}}$, which is expected for our galaxies assuming that the intrinsic UV luminosity equals the total emitted infrared luminosity (energy conservation). For the radius we assume $2$ and $4\,{\rm kpc}$, consistent with the sizes of far-IR emission observed for our galaxies ($\sim0.5\arcsec$ at $z=5.5$). 

\begin{figure}
    \centering
    \includegraphics[width=0.45\textwidth]{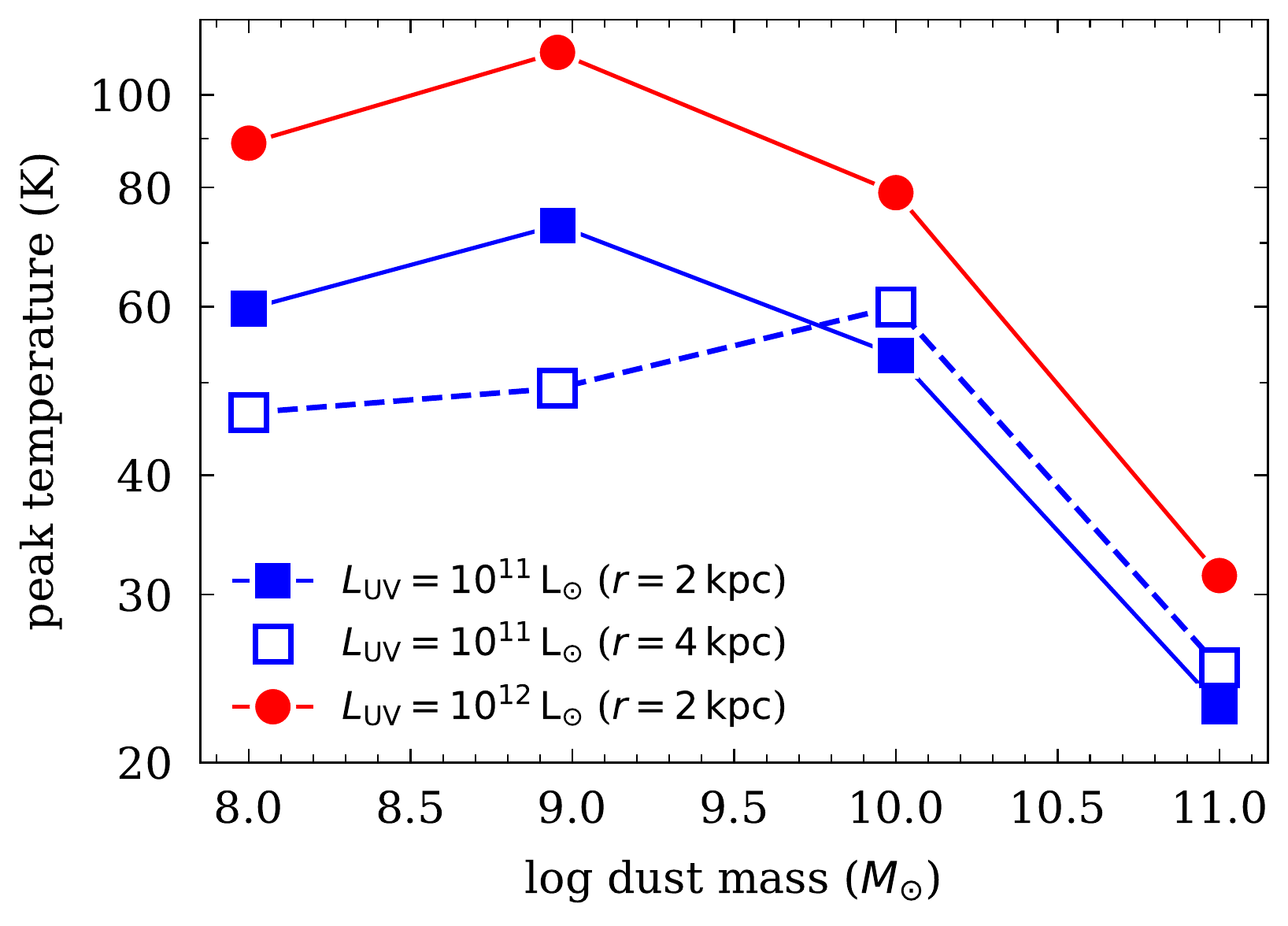}
    \caption{Dependence of peak temperature of emergent light on dust mass and central UV luminosity based on a symmetric shell model (Scoville et al. in prep.). The model assumes two different radii of the spherical dust cloud ($2$ and $4\,{\rm kpc}$) and two intrinsic UV luminosities ($10^{11}$ and $10^{12}\,{L_{\odot}}$) of the embedded source. Due to the fixed size of the dust cloud, the dust volume density and dust opacity increase to the right. The temperature drops at high dust masses (high opacity) because the hotter dust becomes optically thick. The temperature ceases to rise at a dust mass density of $3\times10^{7}\,{\rm M_{\odot}\,kpc^{3}}$ independent of UV luminosity.
    }
    \label{fig:dustmodel}
\end{figure}

Figure~\ref{fig:dustmodel} shows the peak temperature of the spectrum of emergent light computed for our different models as a function of dust mass. Note that dust mass in this case directly corresponds to dust density, hence opacity, as the radius of the cloud is fixed.
This figure shows several trends.
First, the change in peak temperature as a function of UV luminosity is apparent. Specifically, increasing the UV luminosity by an order of magnitude increases the temperature by a factor of $\sim 1.6$. This is expected because $T_{\rm peak} \propto L_{\rm UV}^{1/5}$ for emissivity varying as $\lambda^{-1}$ and optically thin dust at the far-IR peak.
Second, for a given UV luminosity, the peak temperature increases for decreasing opacity (i.e., dust mass or density). This can be explained by the fact that hot dust at the peak of the infrared SED becomes visible to the observer as the opacity drops. At a certain value of opacity, the temperature ceases to rise. For a dust cloud radius of $2\,{\rm kpc}$ ($4\,{\rm kpc}$), this is reached at a dust mass of $10^{9}\,{\rm M_{\odot}}$ ($10^{10}\,{\rm M_{\odot}}$), which translates into an \textit{average} dust mass density of $\bar{\rho}_{d} \sim 3\times 10^{7}\,{\rm M_{\odot}\,kpc^{-3}}$. Note that this number is independent of the intrinsic UV luminosity of the dust-enshrouded source.

Taking the output of this model at face value, the general increase of dust peak temperature with increasing redshift (at roughly fixed total infrared luminosity) can be explained by a decreasing dust opacity, which causes hot dust at short wavelengths to become optically thin and therefore visible to ALMA. In fact, several observations point in this direction. For example, the blue UV continuum slopes of galaxies in the early universe suggest that UV light is less attenuated by dust \citep[e.g.,][]{BOUWENS14}. At the same time, the fraction of dust-obscured star formation decreases significantly at $z>4$ \citep{ALPINE_FUDAMOTO20}.
The current lack of galaxies observed with very hot temperatures at $z>5$ (that would be expected by the trends found at $z<4$) can also be motivated by our model. As the dust opacity (i.e., dust mass density) continues do decrease at higher redshifts, the hot dust becomes optically thin and its temperature ceases to rise (Figure~\ref{fig:opacity}).
In our model, this happens at an average dust mass density of $3\times 10^{7}\,{\rm M_{\odot}\,kpc^{-3}}$. This is indeed in similar to what is expected for our galaxies: Assuming an average molecular gas mass of $3\times10^{10}\,{\rm M_\odot}$ (Section~\ref{sec:Mism}), a gas-to-dust ratio of $100$, and an average size of $2\,{\rm kpc}$, we estimate a dust mass density of $\sim10^{7}\,{\rm M_{\odot}\,kpc^{-3}}$.

\begin{figure}
    \centering
    \includegraphics[width=0.45\textwidth]{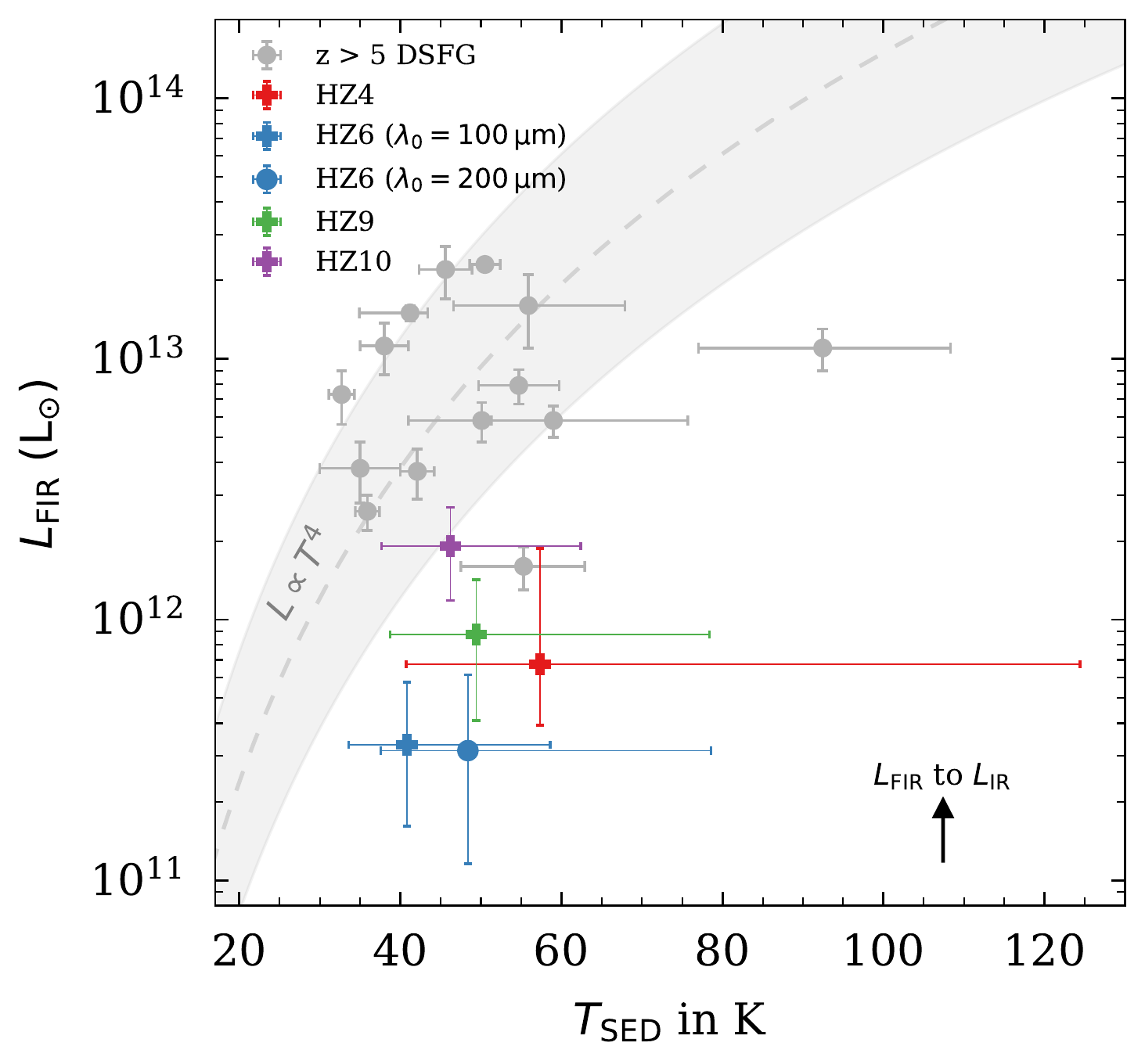}
    \caption{Comparison of SED temperature and \textit{far-IR} luminosity ($\LFIR$) to a sample of DSFGs at $z>5$ from \citet[][]{RIECHERS20}. The luminosity is measured consistently between $42.5\,{\rm \mu m}$ and $122.5\,{\rm \mu m}$ (the arrow shows the difference to $\LIR$). Our galaxies fall off the $L \propto T^{4}$ relation (here normalised median luminosity and temperature of the DSFGs with $0.5\,{\rm dex}$ range indicated). For a given luminosity, they are warmer than expected from this relation. This offset may be caused by different dust abundances, dust surface densities, or dust opacities between DSFGs and normal galaxies.
    }
    \label{fig:riechers}
\end{figure}

\subsection{Comparison with Dusty Star-Forming Galaxies at $z>5$}\label{sec:dsfgs}

In Figure~\ref{fig:riechers}, we compare the far-IR luminosity and SED temperatures of our galaxies to a compilation of infrared luminous dusty star-forming galaxies (DSFGs) at $z>5$ from \citet[][]{RIECHERS20} from the \textit{CO Luminosity Density at High Redshift} survey \citep[COLDz,][]{PAVESI18b,RIECHERS19}.
For a fair comparison, we show SED dust temperatures and far-IR luminosities ($\LFIR$\footnote{For our sample, we find that $\log(\LFIR/{\rm L_{\odot}}) \simeq \log(\LIR/{\rm L_{\odot}}) - 0.2$.}).
One would expect that for an increasing far-IR luminosity, the dust temperature increases (c.f. Figure~\ref{fig:dustmodel}). This is indicated by the $L\propto T^{4}$ relation (optically thick case) normalised to the median of the DSFGs,
\begin{equation}
    \LFIR = 9.4 \times 10^{12} \, \left( \frac{T_{\rm SED}}{50.1} \right)^4\,\,L_{\odot}.
\end{equation}
However, our galaxies seem to be significantly warmer than predicted by this relation, or, for a given temperature their far-IR luminosity is too faint. Formulated in a different way, over 2.5 orders of magnitudes in infrared luminosity, the peak dust temperature is constant, which is in direct contradiction to what is found in the local universe \citep[e.g.,][]{MAGNELLI14}. Since the galaxies are at similar redshifts, this indicates a fundamental difference in the dust properties of in the two samples.
Capitalising on the previous sections and our analytical model, a higher dust abundance and/or dust surface density in the DSFGs would explain the observed differences. Furthermore, as mentioned above and in \citet[][]{FAISST17b}, metallicity (likely connected to dust opacity) has a strong impact on the SED and peak dust temperature. A lower metallicity in our galaxies compared to the DSFGs would increase their temperature at a fixed far-IR luminosity and push them off the $L\propto T^4$ relation.

\subsection{A Final Note on Implication on $\LIR$ Measurements}\label{sec:implicationLIR}

The evolution of the shape of the infrared SED with redshift has important consequences on the measurement of the total infrared luminosity. This quantity is important in several ways, for example, for the computation of dust masses and total star formation rates or the dust properties of high-redshift galaxies via the study of the \IRXB~relation.
For surveys such as \textit{ALPINE}, which target large numbers of main-sequence high-redshift galaxies, only one far-IR data point at $150\,{\rm \mu m}$ exists per galaxy. The above quantities therefore depend strongly on the assumed shape of the infrared SED ($\beta_{\rm d}$, $\alpha$, and temperature).
Using the 3-band constraints on the infrared SEDs of our four galaxies, we can test previous measurements of the total infrared luminosity that are based on only the $150\,{\rm \mu m}$ continuum data point.

We derive total infrared luminosities between $\log(\LIR) = 11.7-12.5$ for our galaxies (Table~\ref{tab:fits}).
Previously obtained luminosities by \citet[][]{CAPAK15}, based on $150\,{\rm \mu m}$ continuum only, also assumed Equation~\ref{eq:firsed1}, however, a lower temperature prior ($T_{\rm SED} = 25-45\,{\rm K}$ or $T_{\rm peak} = 20-30\,{\rm K}$) and $\lambda_0 = 200\,{\rm \mu m}$, but consistent emissivity range ($\beta_{\rm d} = 1.2-2.0$). With these assumptions, $\LIR$ would be underestimated consistently by $0.3-0.6\,{\rm dex}$ (factors $2-4$). 
In \citet[][]{ALPINE_BETHERMIN19}, an average infrared SED created from stacked photometry of COSMOS galaxies between $4 < z < 6$ is normalised to the $150\,{\rm \mu m}$ data points of the \textit{ALPINE} galaxies to derive their total luminosities. This approach leads to consistent total infrared luminosities with ours within less than $0.2\,{\rm \dex}$ ($<60\%$ difference). This result is also reflected in the good agreement of $T_{\rm peak}$ between their average infrared SED and our best fits (c.f., Figure~\ref{fig:tempevolution}).
This comparison shows that (at least statistically) the total infrared luminosities of the \textit{ALPINE} sample derived in \citet[][]{ALPINE_BETHERMIN19} are reasonable and highlights the importance of temperature assumptions in deriving this quantity.

\section{Conclusions}
\label{sec:end}

We have acquired ALMA Band 8 data for four galaxies at $z\sim5.5$ to put improved constraints on their infrared SEDs, specifically their peak dust temperatures, total infrared luminosities, and molecular gas masses. The continuum measurements at a rest-frame wavelength of $\sim110\,{\rm \mu m}$ are blueward of other measurements from the literature in Band 6 ($\sim 200\,{\rm \mu m}$) and Band 7 ($\sim 150\,{\rm \mu m}$) and therefore extend the baseline towards the peak of infrared emission. The infrared SEDs are fit using a modified black body with mid-IR power law. The peak temperature is derived using Wien's law and the molecular gas masses are measured using the extrapolated $850\,{\rm \mu m}$ continuum emission. The measurement of the latter benefits from our so far strongest constraints on the dust emissivity index $\beta_{\rm d}$ at these high redshifts.
In the following, we summarise our findings:

\begin{itemize}{}

\item The best-fit peak temperatures range at $30-43\,{\rm K}$ (median of $38\,{\rm K}$, Figure~\ref{fig:temperaturetraces}). These temperatures are warmer compared to average local galaxies but $5-10\,{\rm K}$ lower than what would be predicted from trends at $z<4$ at similar infrared luminosities. Our measurements are consistent with the most recent hydrodynamical zoom-in simulations, as well as measurements at $z>6$.

\item We find dust emissivity indices ($\beta_d$) between $1.6$ and $2.4$ with a median of $2.0$ (Figure~\ref{fig:temperaturetraces}) for our galaxies, consistent with measurements at lower redshifts.

\item Our new Band 8 data suggest that the emission between rest-frame $100-200\,{\rm \mu m}$ is optically thin (i.e., can be fit with $\lambda_0~=~100\,{\rm \mu m}$) for three of our galaxies (Figure~\ref{fig:opacity}). An exception is \textit{HZ6}, which is a gravitationally interacting three-component major merger and can be fit with optically thick emission below $200\,{\rm \mu m}$ ($\lambda_0=200\,{\rm \mu m}$).

\item The molecular gas masses range between $10^{10}$ and $10^{11}\,{\rm M_{\odot}}$, corresponding to molecular gas fractions between $30\%$ and $80\%$ (Figure~\ref{fig:gasmass}). They are in good agreement with the difference between dynamical and stellar masses. From this, we expect gas depletion time scales of $100-220\,{\rm Myrs}$ in good agreement with the expected decrease of depletion time with redshift. A comparison to gas masses derived from CO($2-1$) emission suggests an $\alpha_{\rm CO}$ conversion factor for \textit{HZ6} and \textit{HZ10} similar to our Milky Way (high values as measured in the SMC can be excluded).

\end{itemize}

At $z<4$, several studies find an increase in dust peak temperature at a roughly fixed total infrared luminosity. Our sample and measurements at $z>6$ do not suggest a further increase of temperature beyond $z=5$. The generally higher peak temperatures at $z=5.5$ compared to average local galaxies can be explained by the decreasing dust abundance (or density) at high redshifts. Specifically, as the dust opacity drops, hot dust becomes more optically thin and is visible to the external observer. The lack of dust temperature evolution at $z>5$ can be explained in similar terms. Our model shows that once the dust density falls below a certain value, the emergent peak temperature ceases to rise. Interestingly, this limit is on the same order of magnitude as the average dust mass density expected for our galaxies.

Compared to DSFGs at similar redshifts ($z>5$), our galaxies have warmer temperatures than what would be expected from their (factor of $10$) lower infrared luminosities. This difference could be explained by a larger dust abundance and/or higher metal content of DSFGs and is in agreement with our model predictions. Metallicity measurements with the \textit{James Webb Space Telescope} for these two populations of galaxies will certainly help to identify what causes these differences.

One of the remaining interesting question is the connection between dust and gas. While a decrease of dust abundance or dust density may explain the observed $T_{\rm peak}-z$ evolution, at the same time the observed increase of the gas fraction (and hence dust abundance given a fixed gas-to-dust ratio) with redshift would argue for the opposite. An increasing gas-to-dust ratio with redshift due to a general decrease in metallicity \citep[e.g.,][]{LEROY11}, could resolve this dilemma.

The number of ALMA observations at high redshifts is increasing rapidly as large surveys are becoming more frequent. Our Band 8 data are an important step to constrain better the infrared SEDs of post-reionisation galaxies. They can be used to inform and improve the assumptions that have to be made in order to measure important infrared SED based quantities as well as to test theoretical predictions.
However, our conclusions are currently based on a sample of only four galaxies, which are, due to observing time constraints, among the infrared brightest galaxies at $z\sim5.5$. Larger samples with similar measurements are crucial to advance our understanding.
As shown by the comparison of gas masses derived by the dust-continuum and the CO($2-1$) emission, at least \textit{HZ6} and \textit{HZ10} have similar CO to ${\rm H_2}$ conversion factors to our Milky Way, which suggests metal enriched environments. This is also suggested by the deep absorption features of their rest-frame UV spectra. It is therefore likely that we are missing more metal poor systems, which could be the more common type of galaxies.
Furthermore, our small sample also shows a diversity of galaxies (isolated galaxies, mergers, etc) that links to different dust properties, which should be explored.
Our simple model can motivate certain trends seen in our sample, but to understand in depth the physics driving the observational results, similar observations for larger samples will be necessary in the future.


\section*{Acknowledgements}

This paper makes use of the following ALMA data: 
\path{ADS/JAO.ALMA#2018.1.00348.S},
\path{ADS/JAO.ALMA#2017.1.00428.L},
\path{ADS/JAO.ALMA#2015.1.00388.S},
\path{ADS/JAO.ALMA#2015.1.00928.S},
\path{ADS/JAO.ALMA#2012.1.00523.S}.
ALMA is a partnership of ESO (representing its member states), NSF (USA) and NINS (Japan), together with NRC (Canada), MOST and ASIAA (Taiwan), and KASI (Republic of Korea), in cooperation with the Republic of Chile. The Joint ALMA Observatory is operated by ESO, AUI/NRAO and NAOJ.
The National Radio Astronomy Observatory is a facility of the National Science Foundation operated under cooperative agreement by Associated Universities, Inc.
This work was supported by the Swiss
National Science Foundation through the SNSF Professorship grant 157567 `Galaxy Build-up at Cosmic Dawn'.
D.R. acknowledges support from the National Science Foundation under grant numbers AST-1614213 and AST-1910107. D.R. also acknowledges support from the Alexander von Humboldt Foundation through a Humboldt Research Fellowship for Experienced Researchers.




\bibliographystyle{mnras}
\bibliography{bibli} 







\bsp	
\label{lastpage}
\end{document}